\documentclass[aps,pra,twocolumn,groupedaddress,showpacs,superscriptaddress]{revtex4-1}
\usepackage{graphicx}

\begin{document}
\title{Using Off-diagonal Confinement as a Cooling Method}

\author{V.~G.~Rousseau, K.~Hettiarachchilage, M.~Jarrell, J.~Moreno, D.~E.~Sheehy}
\address{Department of Physics and Astronomy, Louisiana State University, Baton Rouge, Louisiana 70803, USA}

\begin{abstract}
In a recent letter [Phys. Rev. Lett. {\bf 104}, 167201 (2010)] we proposed a new confining method for 
ultracold atoms on optical lattices, which is based on off-diagonal confinement (ODC). This method was 
shown to have distinct advantages over the conventional diagonal confinement (DC), that makes use of a 
trapping potential, such as the existence of pure Mott phases and highly populated condensates. In this 
manuscript we show that the  ODC method can also lead to lower temperatures 
than the DC method for a wide range of control parameters. Using exact diagonalization
we determine this range of parameters for the hard-core case; then we extend our results 
to the soft-core case by performing quantum Monte Carlo (QMC) simulations for both DC and ODC systems at fixed 
temperature, and analyzing the corresponding entropies. We also
propose a method for measuring the entropy in QMC simulations.
\end{abstract}

\maketitle

\section{Introduction}
With recent experimental developments on cold atoms in optical lattices, the interest in the bosonic 
Hubbard model~\cite{Fisher,Batrouni1990} has dramatically increased. This model is characterized by a 
superfluid-to-Mott quantum phase transition for large onsite repulsion and integer values of the density 
of particles. In actual experiments the atoms are confined to prevent them from leaking out of the lattice. 
This is currently achieved by applying a spatially dependent magnetic field~\cite{Greiner}. A parabolic 
potential is added into the Hubbard model~\cite{Batrouni2002} to mimic the effect of the magnetic field. 
Therefore, the resulting model does not exhibit a true superfluid-to-Mott transition, since Mott 
regions always coexist with superfluid regions. This was predicted theoretically~\cite{Batrouni2002}, and later 
confirmed experimentally~\cite{Folling}.

Recently, we have proposed a new confining technique~\cite{RousseauODC} where the atoms are confined via a 
hopping integral that decreases as a function of the distance from the center of the lattice. Since the 
confinement of the particles is due to the hopping or off-diagonal operators, we  called it 
\textit{Off-Diagonal Confinement} (ODC), as opposed to the conventional diagonal confinement (DC) which makes 
use of a parabolic confinement potential that is reflected in the density profile~\cite{Batrouni2002}. 
For large on-site repulsion the ODC model exhibits pure Mott phases at commensurate filling while at other
fillings it exhibits more populated condensates than the DC model. Another advantage of ODC is that simple 
energy measurements can provide insights into the Mott gap, while the presence of the harmonic potential
may renormalize the value of the gap with respect to the uniform case~\cite{Carrasquilla}. 

In this paper, we show that the ODC method can also lead to lower temperatures than the DC method for a wide 
range of parameters. Producing low temperatures in experiments is challenging, especially with fermions for 
which laser cooling is not as efficient as for bosons. In current experiments, fermions are cooled down by 
convection in the presence of cold bosons, leading to Bose-Fermi mixtures~\cite{Ott,Cazalilla,Hebert,Zujev}. 
Achieving lower temperatures for bosonic condensates will therefore result in colder Bose-Fermi mixtures. 
%This motivates further the interest for the ODC method.

This manuscript is organized as follows. In section II we define our model and describe our methods.
The hard-core limit is studied in section III in order to illustrate analytically how ODC produces 
temperatures that are lower than those obtained with DC. This will also serve to benchmark the quantum 
Monte Carlo (QMC) simulations we use for analyzing the general soft-core case. In section IV we present 
the algorithm we use for QMC simulations, and we propose a method for measuring the entropy with this 
algorithm. Results for the soft-core case are presented in section V. Finally we conclude in section VI.

\section{Model and method}

We consider bosons confined to a one-dimensional optical lattice with $L$ sites and lattice constant $a=1$. 
%Thus the (one-dimensional) volume of the system is $V=L$. 
The Hamiltonian takes the form:
\begin{eqnarray}
  \label{Hamiltonian} \nonumber \hat\mathcal H &=& -\sum_{\big\langle i,j\big\rangle}t_{ij}
\Big(a_i^\dagger a_j^{\phantom\dagger}+h.c.\Big)+\frac{U}{2}\sum_i \hat n_i(\hat n_i-1)\\
                                     & & +W\sum_i(i-L/2)^2\hat n_i
\end{eqnarray}
The creation and annihilation operators $a_i^\dagger$ and $a_i^{\phantom\dagger}$ satisfy bosonic 
commutation rules, $\big[a_i^{\phantom\dagger},a_j^{\phantom\dagger}\big]=
\big[a_i^\dagger,a_j^\dagger\big]=0$, $\big[a_i^{\phantom\dagger},a_j^\dagger\big]=\delta_{ij}$,
and $\hat n_i=a_i^\dagger a_i^{\phantom\dagger}$ is the number of bosons on site $i$.
The sum $\sum_{\langle i,j\rangle}$ runs over all distinct pairs of first neighboring sites $i,j$, and $t_{ij}$ 
is the hopping integral between $i$ and $j$. The parameter $U$ is 
the strength of the local on-site interaction, and $W$ describes the 
curvature of the external trapping potential.

In this work we consider the grand-canonical partition function,
\begin{equation}
  \label{PartitionFunction} \mathcal Z=\textrm{Tr }e^{-\beta\big(\hat\mathcal H-\mu\hat\mathcal N\big)},
\end{equation}
where $\displaystyle \beta=\frac{1}{k_B T}$, $k_B$ is the Boltzmann constant and $T$ the temperature.
The chemical potential $\mu$ controls the average number of particles, $N=\langle\hat\mathcal N\rangle$, 
with $\hat\mathcal N=\sum_i\hat n_i$.  The conventional DC model is obtained by setting $t_{ij}=1$ for 
all pairs of first neighboring sites $i,j$, and using $W>0$. For this model the value of $L$ is irrelevant 
as long as it is sufficiently large to contain the whole gas. The ODC model is obtained by setting $W=0$ 
and using a hopping integral $t_{ij}$ that decreases as a function of the distance from the center of the 
lattice, and vanishes at the edges. For this model, $L$ fully determines $t_{ij}$ as described below.

Typically the temperature is not a control parameter in cold-atoms experiments, and 
once laser cooling has been performed, the system has a fixed entropy which
can be considered the control parameter. Then the temperature can be estimated numerically knowing 
the isentropies of the system~\cite{Mahmud}. Therefore, our strategy for determining which of the two 
confining methods can achieve the lowest temperature is based on switching adiabatically 
from DC to ODC, so the entropy is conserved. Then we determine the temperatures $T_{dc}$ and $T_{odc}$ 
of the DC and ODC systems by equating the entropies

We will consider an experiment in which a \textit{fixed} number $N$ of atoms is loaded
into an optical lattice with a DC trap, described by Eq.~(\ref{Hamiltonian}) with parameters 
$t_{ij}=1$, $W=0.008$ (as in Ref.~\cite{Batrouni2002}). We use $L=400$ in
order to ensure the confinement of the whole gas. Then, we adiabatically switch to the  ODC trap by 
slowly varying $t_{ij}$ and $W$ to $t_{ij}=(i+j+1)(2L-i-j-1)/L^2$ with $L=70$, 
and $W=0$ (as in Ref.~\cite{RousseauODC}), keeping $N$ and $U$ the same.

However, in our calculation, it is actually more convenient to control the temperature than the entropy. 
Thus we consider both DC and ODC systems for
a set temperatures $T$, and measure the corresponding entropies, $S_{dc}(T)$ and $S_{odc}(T)$. 
Then, knowing the initial temperature $T_{dc}$, the final temperature
$T_{odc}$ can be extracted graphically by imposing the equality,
$S_{dc}(T_{dc})=S_{odc}(T_{odc})$, as described in the next section.

\section{The hard-core case: Exact analytical results}

The hard-core limit ($U=+\infty$) of the model can be solved analytically. These exact results provide 
a solid benchmark for our study of the general soft-core case in the next section. We follow here the method 
used by Rigol~\cite{Rigol2005}. In the hard-core limit, the $U$ term in (\ref{Hamiltonian}) can be dropped if 
the standard bosonic commutation rules
are replaced by $\big[a_i^{\phantom\dagger},a_j^{\phantom\dagger}\big]=\big[a_i^\dagger,a_j^\dagger\big]=
\big[a_i^{\phantom\dagger},a_j^\dagger\big]=0$
for $i\neq j$, and $a_i^{\phantom\dagger}a_i^\dagger+a_i^\dagger a_i^{\phantom\dagger}=1$, and 
$a_i^{\phantom\dagger2}=a_i^{\dagger2}=0$. With this algebra, the model (\ref{Hamiltonian}) reduces to
\begin{equation}
  \label{HardcoreHamiltonian} \hat\mathcal H=-\sum_{\big\langle i,j\big\rangle}t_{ij}\Big(a_i^\dagger 
a_j^{\phantom\dagger}+h.c.\Big)+W\sum_i(i-L/2)^2\hat n_i,
\end{equation}
which describes hard-core bosons. By performing a Jordan-Wigner transformation, 
the hard-core creation and annihilation operators can be mapped onto
fermionic creation and annihilation operators, $f_i^\dagger$ and $f_i^{\phantom\dagger}$,
\begin{equation}
  a_j^\dagger=f_j^\dagger \prod_{q=1}^{j-1}e^{i\pi f_q^\dagger f_q^{\phantom\dagger}},\quad a_j^{\phantom\dagger}=
\prod_{q=1}^{j-1}e^{-i\pi f_q^\dagger f_q^{\phantom\dagger}}f_j^{\phantom\dagger},
\end{equation}
which satisfy the usual fermionic anticommutation rules, $\big\lbrace f_i^{\phantom\dagger},f_j^{\phantom\dagger}
\big\rbrace=\big\lbrace f_i^\dagger,f_j^\dagger\big\rbrace=0$,
$\big\lbrace f_i^{\phantom\dagger},f_j^\dagger\big\rbrace=\delta_{ij}$.
This leads to a model that describes free spinless fermions,
\begin{equation}
  \label{FermionicHamiltonian} \hat\mathcal H=-\sum_{\big\langle i,j\big\rangle}t_{ij}
\Big(f_i^\dagger f_j^{\phantom\dagger}+h.c.\Big)+W\sum_i(i-L/2)^2\hat n_i,
\end{equation}
where $\hat n_i=f_i^\dagger f_i^{\phantom\dagger}$ represents the number of fermions on site $i$. Because 
the model (\ref{FermionicHamiltonian}) is a quadratic form of $f_i^\dagger$ and $f_i^{\phantom\dagger}$,
it can be solved by a simple numerical diagonalization of the $L\times L$ matrix. Denoting by $\epsilon_k$ 
with $k\in[1,L]$ the eigenvalues of this matrix, the partition function (\ref{PartitionFunction}) takes the
form
\begin{equation}
  \label{PartitionFunction2} \mathcal Z=\prod_{k=1}^L\Big(1+e^{-\beta(\epsilon_k-\mu)}\Big).
\end{equation}
The entropy is defined as $S=-k_B\textrm{Tr }\mathcal D\ln\mathcal D$ with the density matrix
$\displaystyle\mathcal D=\frac{1}{\mathcal Z}e^{-\beta(\hat\mathcal H-\mu\hat\mathcal N)}$. Working in a system of units where 
the Boltzmann constant $k_B=1$ and using the properties of the density matrix, it follows that $S=\ln\mathcal 
Z+\beta\big\langle\hat\mathcal H\big\rangle-\beta\mu\big\langle\hat\mathcal N\big\rangle$.
Substituting $\displaystyle\big\langle\hat\mathcal H\big\rangle-\mu\big\langle\hat\mathcal N\big\rangle=
-\frac{\partial}{\partial\beta}\ln\mathcal Z$ and using expression (\ref{PartitionFunction2}) for $\mathcal Z$, 
the entropy takes the form
\begin{equation}
  \label{Entropy} S(\beta,\mu)=\sum_{k=1}^L\Big[\ln\Big(1+e^{-\beta(\epsilon_k-\mu)}\Big)+
\frac{\beta(\epsilon_k-\mu)}{e^{\beta(\epsilon_k-\mu)}+1}\Big].
\end{equation}
The average number of particles $N$ is obtained by summing the Fermi-Dirac distribution,
\begin{equation}
  \label{FermiDirac} N(\beta,\mu)=\sum_{k=1}^L\frac{1}{e^{\beta(\epsilon_k-\mu)}+1}.
\end{equation}

\begin{figure}[tb]
  \centerline{\includegraphics[width=0.45\textwidth]{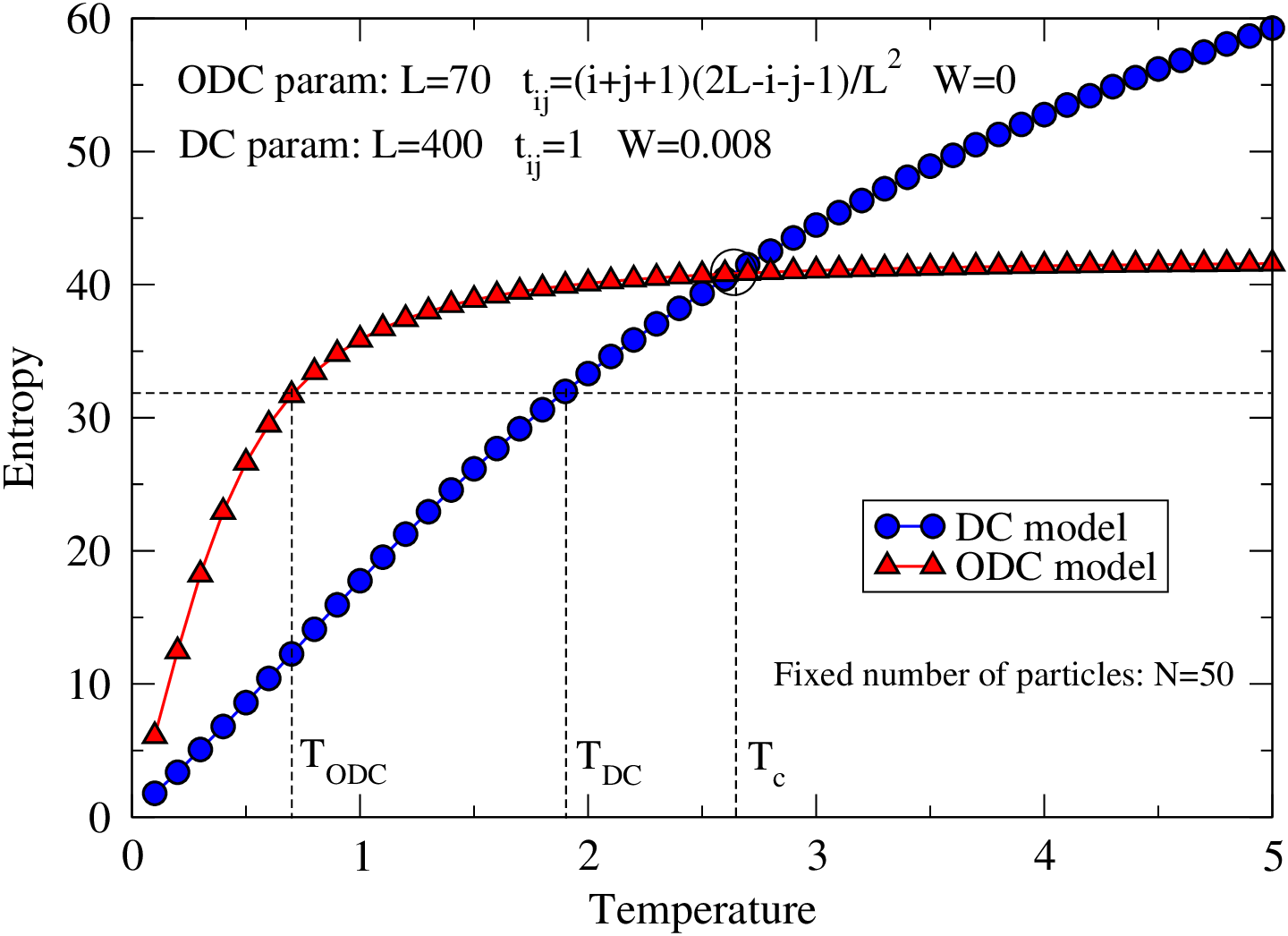}}
  \caption
    {
      (Color online) The entropy as a function of temperature for 50 hard-core bosons, in the DC case (circles) 
and in the ODC case (triangles). There exists a critical temperature $T_c$
      where the two curves cross. If the initial temperature $T_{dc}$ is below $T_c$, then 
the conservation of the entropy when switching adiabatically from
      DC to ODC implies that the final temperature $T_{odc}$ is lower.
    }
  \label{EntropyHardcoreN50}
\end{figure}

Fig.~\ref{EntropyHardcoreN50} shows the entropy~(\ref{Entropy}) as a function of temperature for both DC and 
ODC cases. The chemical potential $\mu$ is adjusted such that the average number of particles~(\ref{FermiDirac})
remains constant ($N=50$). An interesting feature is that the two curves cross at a 
temperature $T_c$, and that below $T_c$ the entropy of the ODC system is greater
than the entropy of the DC system. Thus, if the initial temperature $T_{dc}$ is below $T_c$, then the final 
temperature $T_{odc}$ is lower when switching adiabatically from DC to ODC.

Next we generalize our discussion by calculating, for a fixed number of particles $N$, the critical temperature $T_c$ 
below which the ODC method produces a temperature lower than the DC method when the confinement is switch 
adiabatically. In order to determine $T_c$ for  given  parameters $t^{dc}_{ij}$ and 
$W$ for the conventional DC system, and $t^{odc}_{ij}$ for  the ODC model, one needs to solve for each
value of $N$ a system of three coupled non-linear equations,
\begin{equation}
\label{System}
  \left\lbrace\begin{array}{l}
    S_{odc}(\beta_c,\mu_{odc})=S_{dc}(\beta_c,\mu_{dc}) \\
    N_{odc}(\beta_c,\mu_{odc})=N \\
    N_{dc}(\beta_c,\mu_{dc})=N
  \end{array}\right.
\end{equation}
where $S_{odc}$ ($S_{dc}$) is given by Eq.~(\ref{Entropy}) and $N_{odc}$ ($N_{dc}$) is given 
by Eq.~(\ref{FermiDirac}), with $\mu=\mu_{odc}$ ($\mu=\mu_{dc}$), and $\beta=\beta_c$.
The first equation corresponds to the conservation of the entropy when switching from DC to ODC, and the 
two others correspond to the
conservation of the number of particles. Solving this system of equations determines the critical inverse 
temperature $\beta_c=1/T_c$, and the chemical potentials $\mu_{odc}$ and $\mu_{dc}$ that give
the desired number of particles $N$.

For this purpose, we define an \textit{error} function:
\begin{eqnarray}
  \nonumber \mathcal E(\beta_c,\mu_{odc},\mu_{dc}) &=& \big(S_{odc}(\beta_c,\mu_{odc})-
S_{dc}(\beta_c,\mu_{dc})\big)^2\\
  \nonumber                                        & & +\big(N_{odc}(\beta_c,\mu_{odc})-N\big)^2\\
                                                   & & +\big(N_{dc}(\beta_c,\mu_{dc})-N\big)^2.
\end{eqnarray}
By construction, the solution of Eq.~(\ref{System}) minimizes this error function.
Starting with an initial guess for $\beta_c$, $\mu_{odc}$, and $\mu_{dc}$, we calculate the error 
$\mathcal E(\beta_c,\mu_{odc},\mu_{dc})$ and its gradient 
$\vec\nabla\mathcal E=(\partial\mathcal E/\partial\beta_c,\partial\mathcal 
E/\partial\mu_{odc},\partial\mathcal E/\partial\mu_{dc})$.
Writing the initial guess as a vector, $\vec r=(\beta_c,\mu_{odc},\mu_{dc})$, we perform a correction 
$\Delta\vec r$ by following the opposite direction of the gradient $\vec\nabla\mathcal E$.
Then we iterate until convergence.

\begin{figure}[tb]
  \centerline{\includegraphics[width=0.45\textwidth]{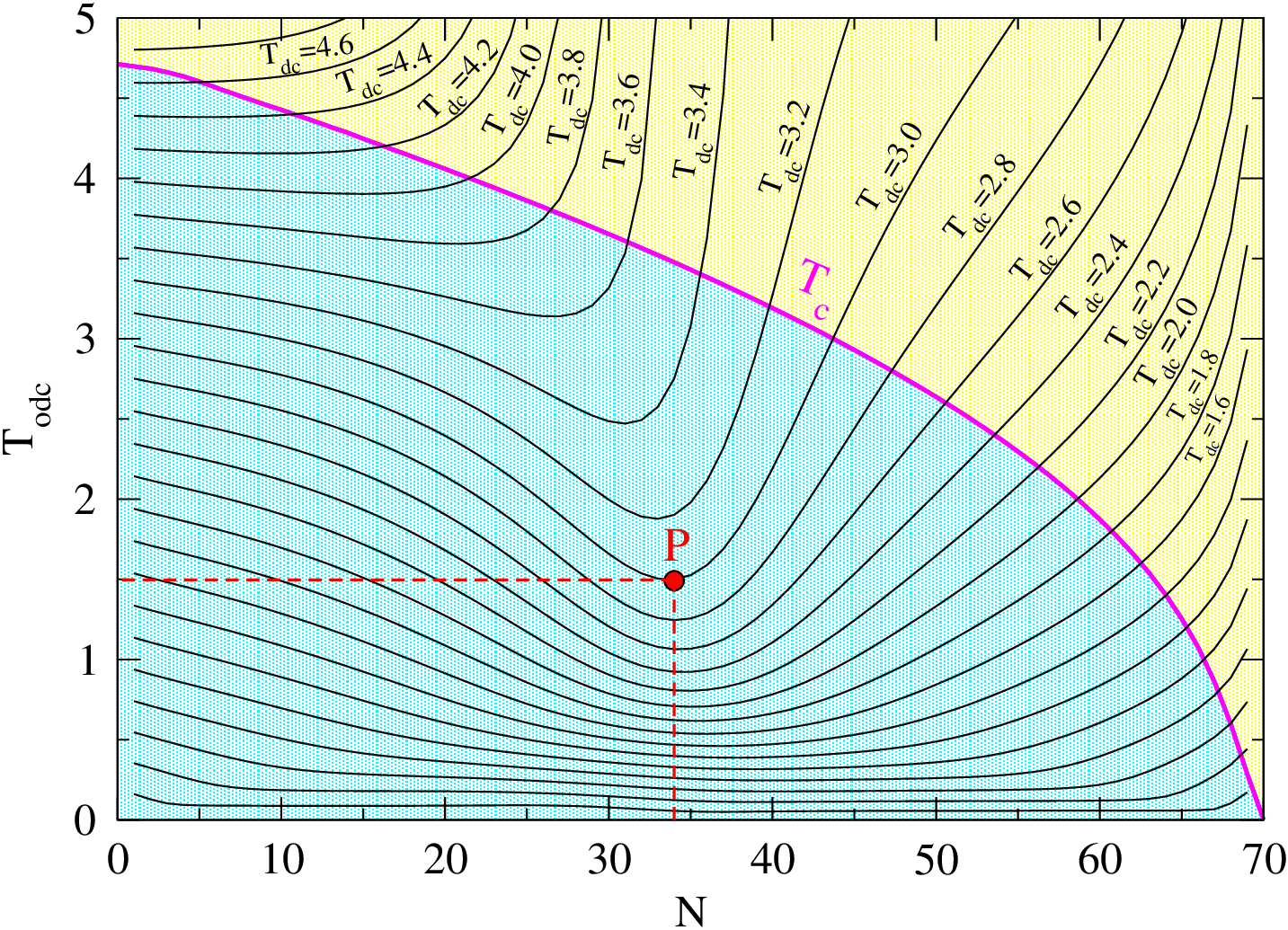}}
  \caption
    {
      (Color online) The critical temperature $T_c$ and the DC isotherms as functions of the number of 
particles $N$. At $T=T_c$, there is no change in temperature when switching adiabatically from
      DC to ODC. Below $T_c$ (blue region), the ODC method gives a temperature $T_{odc}$ that is lower than 
the temperature $T_{dc}$ obtained with the DC method. Above $T_c$ (yellow region),
      it is the DC method that gives the lowest temperature. For example, the point $P$ corresponds to a system 
with 34 particles, an initial temperature $T_{dc}=3$, and a final temperature
      $T_{odc}\approx 1.5$.
    }
  \label{TcVsN}
\end{figure}

Fig.~\ref{TcVsN} shows the critical temperature $T_c$ and the DC isotherms as functions of $N$. For a 
given number of particles and an initial temperature $T=T_c(N)$, the ODC and DC systems have the same 
temperature $T_{odc}=T_{dc}$ when the confinement is switch adiabatically.
Below (above) $T_c$, the ODC system has a temperature $T_{odc}$ that is lower (higher) than the temperature 
$T_{dc}$ of the DC system. The point $P$ illustrates how the figure should
be read: For a system with 34 particles and an initial DC temperature $T_{dc}=3$, the final ODC temperature 
is $T_{odc}\approx 1.5$. Note that $T_c$ vanishes when N=L=70.  The resulting Mott phase found in the ODC case
always has lower entropy than the mixed phases found in the DC case.  This will be discussed in greater detail 
in the next section.

\section{Quantum Monte Carlo algorithm and the entropy}
For the treatment of soft-core interactions, we perform QMC simulations using the Stochastic Green 
Function (SGF) algorithm~\cite{SGF} with tunable directionality~\cite{DirectedSGF}. Although this
algorithm was developed for the canonical ensemble, a trivial extension~\cite{Wolak2010} 
allows us to simulate the grand-canonical ensemble.  We propose a new method to measure the 
entropy by taking advantage of the grand-canonical ensemble. Our thermodynamic control parameters are the temperature 
$T$, the volume $V$ (number of sites $L$), and the chemical potential $\mu$.
\begin{figure}[tb]
  \centerline{\includegraphics[width=0.45\textwidth]{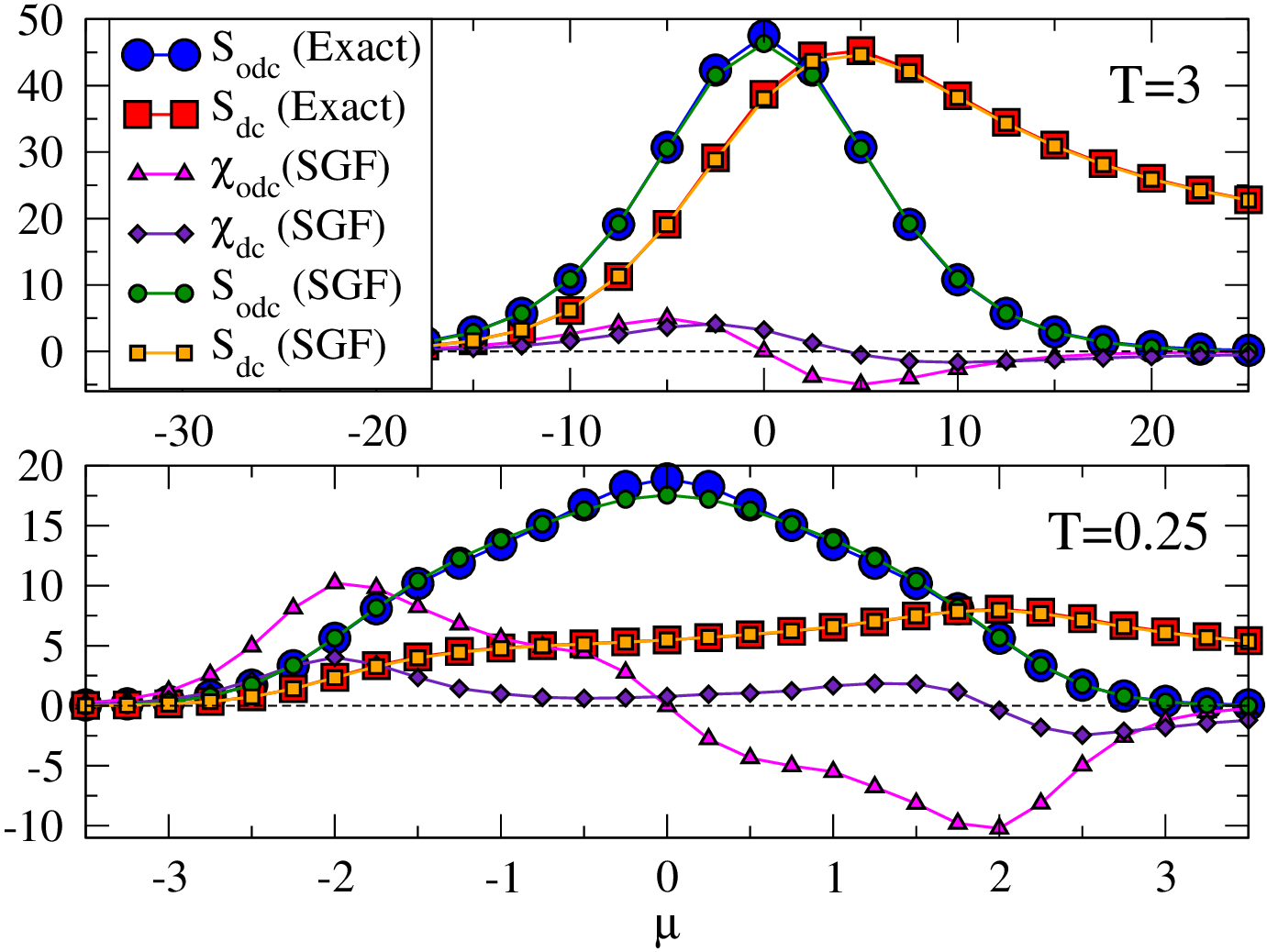}}
  \caption
    {
      (Color online) The entropy $S$ and the thermal susceptibility $\chi$ in the hard-core case. Comparison 
between results obtained with exact diagonalization using Eq.~(\ref{Entropy}) and QMC results with 
the SGF algorithm and Eq.~(\ref{EntropyQMC}). Two different temperatures are considered, $T=3$ and $T=0.25$, 
for both DC and ODC systems.
    }
  \label{Comparison}
\end{figure}
Unlike the analytical hard-core case, a direct measurement of the entropy is not possible with a single QMC 
simulation because the value of $\mathcal Z$ is unknown. However it is still possible to evaluate the entropy
with a set of QMC simulations. For this purpose, we define the \textit{thermal susceptibility} 
$\chi_{th}$ by the response of the number of particles $N$ to an infinitesimal change
of the temperature $T$:
\begin{equation}
  \label{ThermalSusceptibility} \chi_{th}(T,V,\mu)=\frac{\partial N}{\partial T}\Big|_{V,\mu}
\end{equation}
By substituting $N=\frac{1}{\mathcal Z}\textrm{Tr }\hat\mathcal N e^{-\beta(\hat\mathcal H-\mu\hat\mathcal N)}$ in 
expression~(\ref{ThermalSusceptibility}),
we get an expression for the thermal susceptibility that can be directly measured in our simulations:
\begin{equation}
  \label{ThermalSusceptibilityQMC} \chi_{th}=\beta^2\Big[\big\langle\hat\mathcal N
\big(\hat\mathcal H-\mu\hat\mathcal N\big)\big\rangle-\big\langle\hat\mathcal N\big\rangle\big\langle
\big(\hat\mathcal H-\mu\hat\mathcal N\big)\big\rangle\Big]
\end{equation}
Considering the energy $E=\langle\hat\mathcal H\rangle$ and the associated differential $dE=TdS-PdV+\mu dN$, 
where the pressure $P$ is defined as $\displaystyle P=-\frac{\partial E}{\partial V}\Big|_{S,N}$, and performing 
a Legendre transformation over the variables $S$ and $N$, we can define the grand-canonical potential $\Omega$
that depends only on our natural variables, $\Omega(T,V,\mu)=E-TS-\mu N=-PV$. Its differential takes the form
\begin{equation}
  d\Omega=-SdT-PdV-Nd\mu.
\end{equation}
We can then extract a useful Maxwell relation,
\begin{equation}
  \frac{\partial S}{\partial\mu}\Big|_{V,T}=\frac{\partial N}{\partial T}\Big|_{V,\mu},
\end{equation}
so the entropy can be easily obtained by integrating the thermal susceptibility over the chemical potential and 
keeping the temperature and the volume constant,
\begin{equation}
  \label{EntropyQMC} S(T,V,\mu)=\int_{\mu_0}^{\mu}\chi_{th}(T,V,\mu')d\mu',
\end{equation}
where $\mu_0$ is the critical value of the chemical potential below which the average number of particles $N$ 
and the thermal susceptibility $\chi_{th}$ are vanishing.

\begin{figure}[tb]
  \centerline{\includegraphics[width=0.45\textwidth]{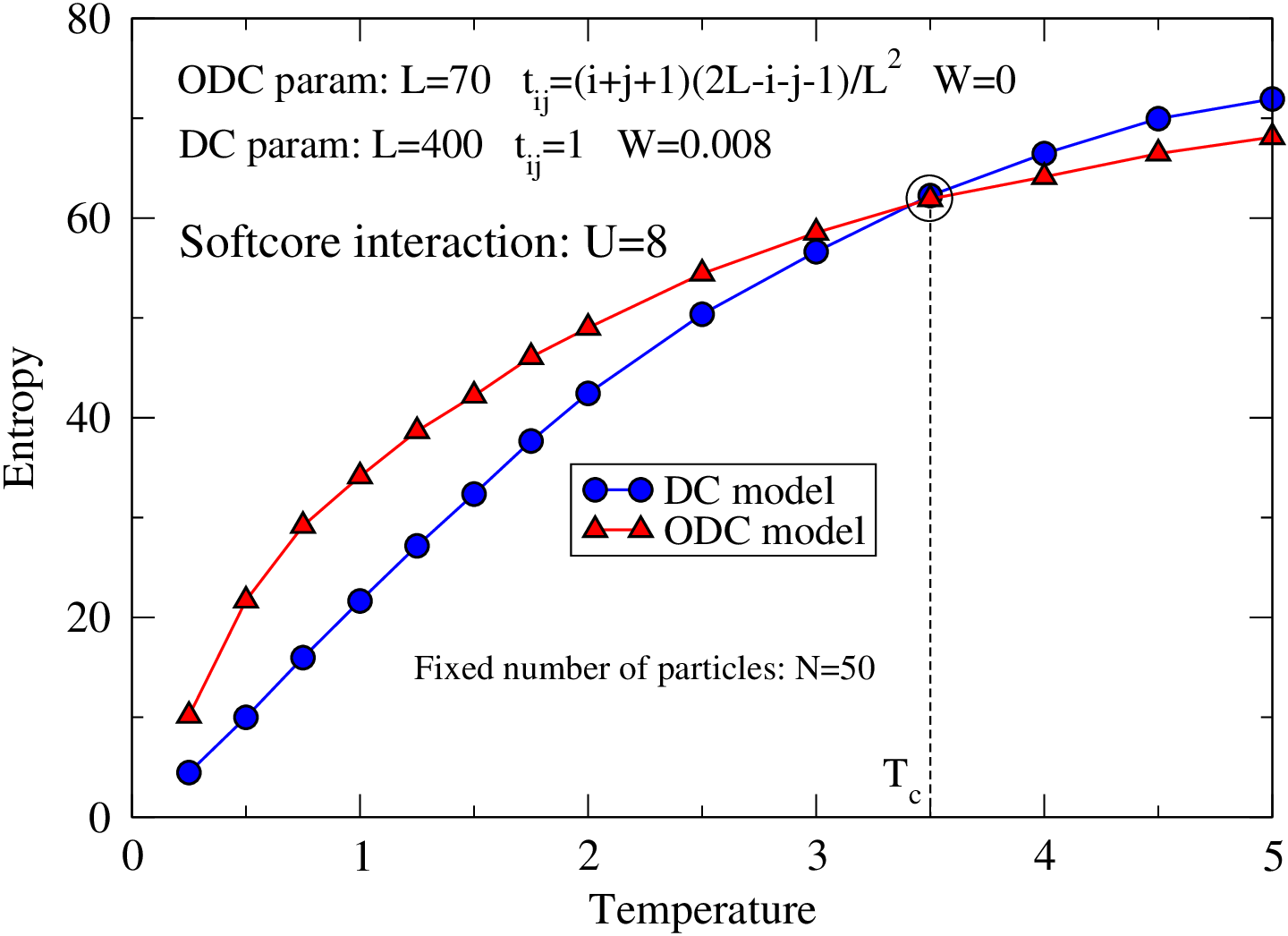}}
  \caption
    {
      (Color online) The entropy as a function of temperature for 50 soft-core bosons with $U=8$, in the DC 
case (circles) and in the ODC case (triangles). As in the hard-core case (Fig.\ref{EntropyHardcoreN50}), 
there exists a critical temperature $T_c$ below which the entropy of the 
ODC system is higher than the entropy of the DC system, thus making the ODC method
      more efficient than the DC method for producing low temperatures.
    }
  \label{EntropySoftcoreN50}
\end{figure}

In order to check the reliability of Eq.~(\ref{EntropyQMC}), we show on Fig.~\ref{Comparison} a comparison of the 
entropy of the hard-core case obtained with the SGF algorithm by
integrating the thermal susceptibility (\ref{ThermalSusceptibilityQMC}), and
the entropy computed with Eq.~(\ref{Entropy}). The agreement is good for 
both DC and ODC systems at high ($T=3$) and low temperatures ($T=0.25$).
\begin{figure}[tb]
  \centerline{\includegraphics[width=0.45\textwidth]{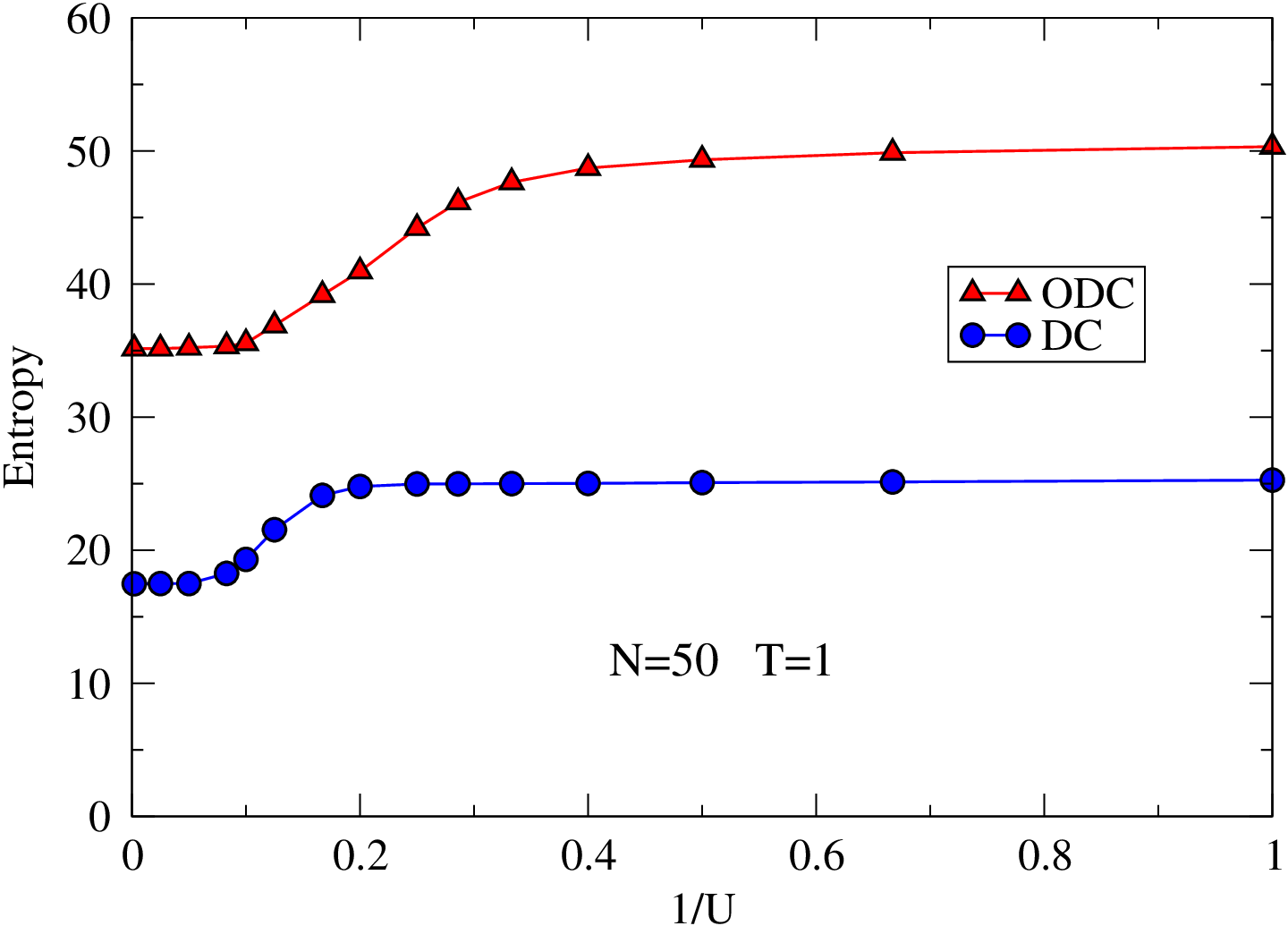}}
  \caption
    {
      (Color online) The entropy as a function of the inverse onsite interaction $1/U$ for 50 soft-core bosons 
at $T=1$, in the DC (circles) and ODC cases (triangles). The entropy of the ODC system remains greater than the 
entropy of the DC system for any value of the onsite repulsion $U$,  showing that our suggested cooling method 
works deep into the soft-core case.
    }
  \label{EntropySoftcoreN50VsU}
\end{figure}

We now release the hard-core constraint and set the onsite repulsion $U=8$. Fig.~\ref{EntropySoftcoreN50} 
shows the entropies for the DC and the ODC models as functions of
temperature for $N=50$. The curves differ from the hard-core case only quantitatively, not 
qualitatively, showing that the method of cooling by switching from DC to ODC
still works. Moreover, one notices that the critical temperature $T_c\approx3.5$ is higher than in the hard-core 
case ($T_c\approx 2.65$) which makes easier to access the regime where
ODC is more efficient than DC.

Further, we extend our soft-core results to different values of the onsite repulsion. 
Fig.~\ref{EntropySoftcoreN50VsU} shows the entropy for the DC and the ODC models as function of the inverse
onsite repulsion $1/U$ for $N=50$ and $T=1.0$. The curves show that the entropy of the ODC model is above the 
one of the DC model for any value of $U$. Thus, for this filling,
the ODC method produce temperatures lower than the DC method for any value of the onsite repulsion. 
When $U$ is large, the results match with those obtained in the preceding
section for the hard-core case (Fig.~\ref{EntropyHardcoreN50}).

\begin{figure}[t]
  \centerline{\includegraphics[width=0.45\textwidth]{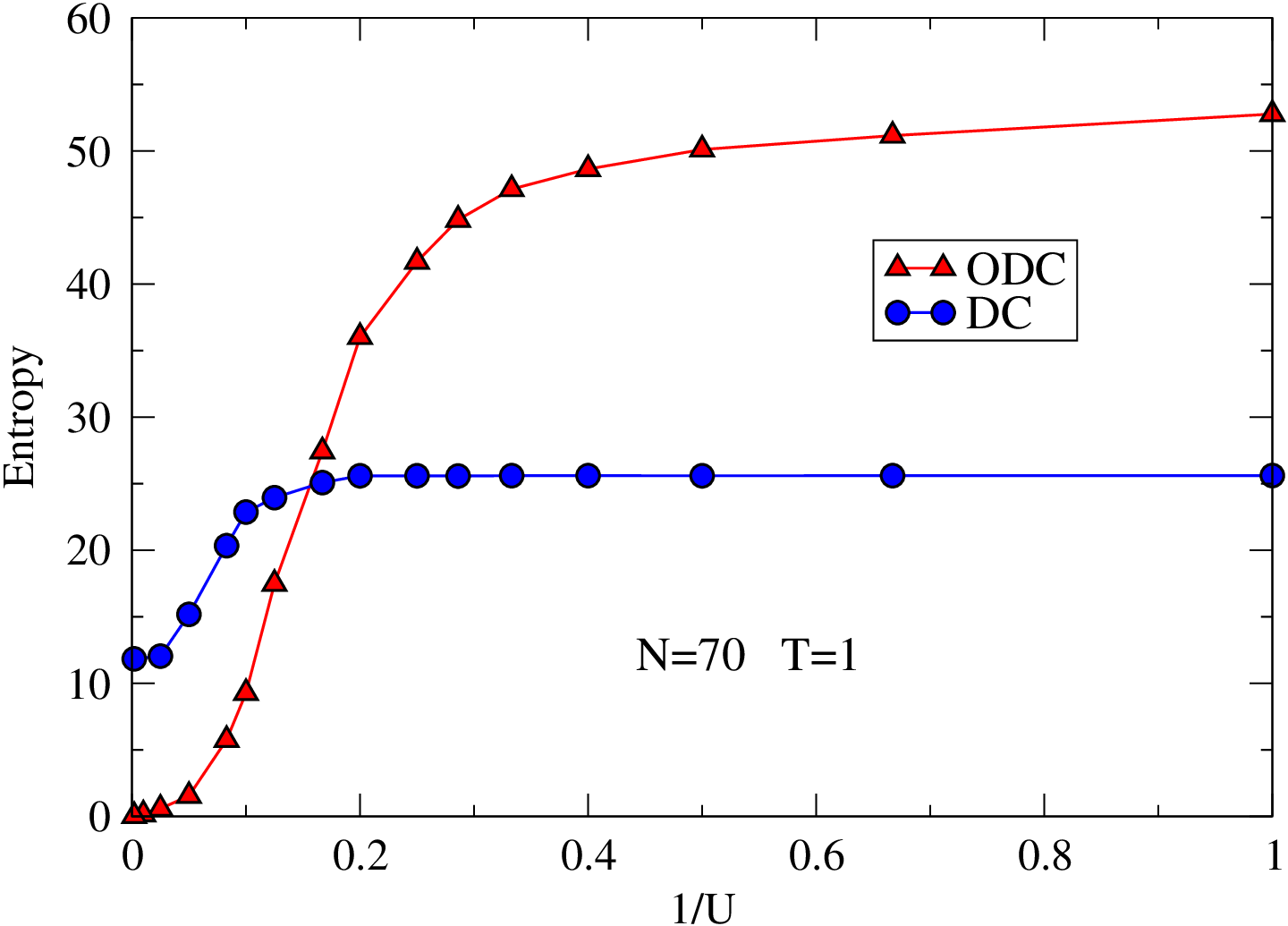}}
  \caption
    {
      (Color online) The entropy as a function of the inverse onsite interaction $1/U$ for 70 soft-core bosons 
at $T=1$, in the DC case (circles) and in the ODC case (triangles). For the ODC case the filling is commensurate 
and the system forms a pure Mott phase that cannot carry any entropy. Thus ODC cannot be used as a cooling method 
for commensurate fillings.
    }
  \label{EntropySoftcoreN70VsU}
\end{figure}

However, the situation is different for $N=70$ as Fig.~\ref{EntropySoftcoreN70VsU} illustrates. At this integer 
filling, the entropy of the ODC model, which vanishes in the large $U$ limit, intersects the curve for the DC model. 
In this regime, the ODC model exhibits a pure Mott phase, hence with zero entropy. However 
the phase of the DC model has Mott regions coexisting with superfluid
regions, so the entropy remains finite. Thus, ODC cannot be used to cool the system in this region.

Concerning the experimental realization of our model, a holographic technique recently developed \cite{Greiner2}
can be used to build the optical lattice with off-diagonal confinement.
Using this method, an off-diagonal trap can be superposed to an existing diagonal trap. Then the diagonal trap
can be turned off. The switching between the two traps can be in principle very fast, however the
technical details of how this will work go beyond the scope of the present manuscript and must be
developed by experimentalists. Nevertheless, a qualitative analysis reveals that three time scales must be
considered. The time scale $\tau_m$ of the model system or roughly $\tau_m=1/t$ (in units where $\hbar=1$),
the time scale of the experiment $\tau_e$, and the time scale $\tau_c$ which describes the coupling of the
model system to its environment which includes the effects of the laser heating, evaporation, etc. In our
proposal, it is important that the trap is adiabatically switch on the experimental time scale, but not on the
time scale which describes the coupling of the trap to its environment, so that $\tau_m << \tau_e << \tau_c$.

\section{Conclusion}
In this manuscript we propose that the adiabatic switch from the DC to the ODC method 
can produce lower temperatures for a wide range of initial temperatures and system parameters.
In the hard-core limit, we determine the critical temperature $T_c$ for which the two methods have
the same entropy. Below (above) $T_c$ and at constant entropy, the ODC method leads to temperatures that are 
lower (higher) than with the DC method. In order to extend our results to the soft-core case, we propose a 
simple method for evaluating the entropy with QMC, by measuring the thermal susceptibility $\chi_{th}$ in the 
grand-canonical ensemble and integrating it over the chemical potential $\mu$. Then we make use of the SGF 
algorithm~\cite{SGF} with tunable directionality~\cite{DirectedSGF}, and show that the soft-core results are 
qualitatively the same as in the hard-core case.

\begin{acknowledgments}
This work was supported by the National Science Foundation through
OISE-0952300, the TeraGrid resources provided by NICS under grant number TG-DMR100007, the high
performance computational resources provided by the Louisiana Optical Network Initiative (http://www.loni.org),
and the Louisiana Board of Regents, under grant LEQSF (2008-11)-RD-A-10. 
\end{acknowledgments}

\end{document}